\documentclass[12pt]{article}

\usepackage{scicite}
\usepackage{graphicx}

\usepackage{times}

\newenvironment{sciabstract}{%
\begin{quote} \bf}
{\end{quote}}

\newcounter{lastnote}

\title{Detection of the Water Reservoir in a Forming Planetary System}

\author {Michiel R. Hogerheijde,$^{1}$ Edwin A. Bergin,$^{2}$
  Christian Brinch,$^{1}$ L. Ilsedore Cleeves,$^{2}$\\ Jeffrey
  K. J. Fogel,$^{2}$ Geoffrey A. Blake,$^{3}$ Carsten Dominik,$^{4}$\\
  Dariusz C. Lis,$^{5}$ Gary Melnick,$^{6}$ David Neufeld,$^{7}$ Olja
  Pani\'c,$^{8}$\\ John C. Pearson,$^{9}$ Lars Kristensen,$^{1}$ Umut
  A. Y{\i}ld{\i}z,$^{1}$ Ewine F.
  van Dishoeck$^{1,10}$\\
  \\
  \normalsize{$^{1}$Leiden Observatory, Leiden University,}\\
  \normalsize{PO Box 9513, 2300 RA Leiden, The Netherlands}\\
  \normalsize{$^{2}$Department of Astronomy, University of Michigan,
    Ann Arbor MI, USA}\\
  \normalsize{$^{3}$Div. of Geological and Planetary Sciences,
    California Institute of Technology, Pasadena CA, USA}\\
  \normalsize{$^{4}$Astronomical Institute `Anton Pannekoek',
    University of Amsterdam, The Netherlands}\\
  \normalsize{$^{5}$Div. of Physics, Mathematics, and Astronomy,
    California Institute of Technology, Pasadena CA, USA}\\
  \normalsize{$^{6}$Harvard-Smithsonian Center for Astrophysics,
    Cambridge MA, USA}\\
  \normalsize{$^{7}$Dept. of Physics and Astronomy, Johns Hopkins
    University, Baltimore MD, USA}\\ 
  \normalsize{$^{8}$European Southern Observatory, Garching, Germany}\\
  \normalsize{$^{9}$Jet Propulsion Laboratory, California Institute of
    Technology, Pasadena CA, USA}\\
  \normalsize{$^{10}$Max-Planck-Institut f\"ur Extraterrestrische
    Physik, Garching, Germany}\\
  \\
  \normalsize{$^\ast$Accepted for publication in Science.}  }

\date{}

\begin{document} 


\baselineskip24pt


\maketitle


\begin{sciabstract}
  Icy bodies may have delivered the oceans to the early Earth, yet
  little is known about water in the ice-dominated regions of
  extra-solar planet-forming disks.  The Heterodyne Instrument for the
  Far-Infrared on-board the Herschel Space Observatory has detected
  emission lines from both spin isomers of cold water vapor from the
  disk around the young star TW Hydrae.  This water vapor likely
  originates from ice-coated solids near the disk surface hinting at a
  water ice reservoir equivalent to several thousand Earth oceans in
  mass.  The water's ortho-to-para ratio falls well below that of
  solar system comets, suggesting that comets contain heterogeneous
  ice mixtures collected across the entire solar nebula during the
  early stages of planetary birth.
\end{sciabstract}

Water in the solar nebula is thought to have been frozen out onto dust
grains outside $\sim 3$ astronomical units (AU\cite{aunote})
\cite{hayashi1981}. Stored in icy bodies, this water provided a
reservoir for impact delivery of oceans to the Earth
\cite{matsui1986}. In planet-forming disks, water vapor is thought to
be abundant only in the hot ($>$250 K) inner regions where ice
sublimates and gas-phase chemistry locks up all oxygen in
H$_2$O. Emission from hot ($>$250 K) water has been detected from
several disks around young stars\cite{pontoppidan2010,carr2011}. In
the cold ($\sim$20 K) outer disk water vapor freezes out, evidenced by
spectral features of water ice in a few
disks\cite{terada2007,honda2009}.  However, (inter)stellar ultraviolet
radiation penetrating the upper disk layers desorbs a small fraction
of water ice molecules back into the gas phase\cite{dominik2005},
suggesting that cold ($<$100 K) water vapor exists throughout the
radial extent of the disk. The detection of this water vapor would
signal the presence of a hidden ice reservoir.

We report detection of ground-state rotational emission lines of both
spin isomers of water ($J_{K_AK_C}$ $1_{10}$--$1_{01}$ from
ortho-H$_2$O and $1_{11}$--$0_{00}$ from para-H$_2$O) from the disk
around the pre-main-sequence star TW~Hydrae (TW~Hya) using the
Heterodyne Instrument for the Far-Infrared (HIFI)
spectrometer\cite{degraauw2010} on-board the Herschel Space
Observatory\cite{pilbratt2010} (Fig. 1)\cite{waternote,somnote}.  TW
Hya is a 0.6 M$_\odot$ (solar mass), 10-million-year old T~Tauri
star\cite{webb1999} $53.7\pm 6.2$ pc away from Earth. Its 196 AU
radius disk is the closest protoplanetary disk to Earth with strong
gas emission lines.  The disk's mass is estimated at $2\times
10^{-4}$ to $6\times 10^{-4}$ M$_\odot$ in dust and, using different
tracers and assumptions, between $4\times 10^{-5}$ and 0.06 M$_\odot$
in gas \cite{kastner1997,calvet2002,gorti2011}.  The velocity widths
of the H$_2$O lines (0.96 to 1.17 km~s$^{-1}$) (table S1) exceed by
$\sim$40\% those of cold CO \cite{kastner1997}.  These correspond to
CO emission from the full 196 AU radius rotating disk inclined at
$\sim$7$^\circ$ with only little ($<$65 m~s$^{-1}$)
turbulence\cite{hughes2011}. The wider H$_2$O lines suggest that the
water emission extends to $\sim$115 AU where the gas orbits the star
at higher velocities compared with 196 AU.

To quantify the amount of water vapor traced by the detected lines, we
performed detailed simulations of the water chemistry and line
formation using a realistic disk model matching previous observations
\cite{thi2010,somnote}.  We adopted a conservatively low dust mass of
$1.9\times 10^{-4}$ M$_\odot$ and, using a standard gas-to-dust mass
ratio of 100, a gas mass of $1.9\times 10^{-2}$ M$_\odot$. We explored
the effects of much lower gas-to-dust ratios.  We followed the
penetration of the stellar ultraviolet and x-ray radiation into the
disk, calculated the resulting photodesorption of water and ensuing
gas-phase chemistry including photodissociation, and solved the
statistical-equilibrium excitation and line formation.  The balance of
photodesorption of water ice and photodissociation of water vapor
results in an equilibrium column of water H$_2$O vapor throughout the
disk (Fig. 2). Consistent with other studies\cite{woitke2009}, we find
a layer of maximum water vapor abundance of $0.5\times 10^{-7}$ to
$2\times 10^{-7}$ relative to H$_2$ at an intermediate height in the
disk. Above this layer, water is photodissociated; below it, little
photodesorption occurs and water is frozen out, with an ice abundance,
set by available oxygen, of 10$^{-4}$ relative to H$_2$.

In our model, the 100- to 196-AU region dominates the line emission,
which exceeds observations in strength by factors of $5.3\pm 0.2$ for
H$_2$O $1_{10}$--$1_{01}$ and $3.3\pm 0.2$ for H$_2$O
$1_{11}$--$0_{00}$.  A lower gas mass does not decrease the line
intensities, if we assume that the water ice, from which the water
vapor derives, formed early in the disk's evolution, before
substantial gas loss occurred, and remains frozen on grains.  The most
plausible explanation involves a difference in the relative location
of small, bare grains regulating the ultraviolet radiative transport
and larger, ice-carrying grains. Differential settling of large grains
relative to small grains moves much of the ice reservoir below the
reach of the ultraviolet radiation, resulting in less water vapor and
weaker lines. Our model matches the observations if only 12\% of the
original ice content remains above this line\cite{meijerink2009}. A
radially increasing degree of settling of icy grains explains the
observed H$_2$O line widths.

The detected water vapor, resulting from photodesorption, implies an
ice reservoir in the giant planet formation zone and beyond. In our
simulations the $7.3\times 10^{21}$ g of detected water vapor
(equivalent to 0.005 times the mass of Earth's oceans) originate
from a total ice reservoir of $9\times 10^{27}$ g (or several
thousands of Earth's oceans) throughout the disk. The size of this
reservoir is tied to the dust mass contained in the disk, for which we
adopt a conservatively low value. Although the ice reservoir is only
observed indirectly, no known mechanism can remove it from the regions
probed by Herschel. Any smaller ice reservoir implies the
corresponding absence of elemental oxygen that efficiently reforms
water ice on the grains.

The detection of both spin isomers of water vapor allows its
ortho-to-para ratio (OPR) to be derived, because our simulations
indicate that the lines are optically thin. An OPR of $0.77\pm 0.07$
matches our observations\cite{somnote}. This value is much lower than
the OPR range of 2 to 3 observed for solar system
comets\cite{bonev2007}. It is common practice to associate the OPR
with a spin temperature $T_{\rm spin}$ at which a Boltzmann
distribution reproduces the ratio of spin isomers. Our derived OPR
corresponds to $T_{\rm spin}$=$13.5\pm0.5$ K, whereas the range for
solar system comets yields a $T_{\rm spin}$ of $>$20 K.

Radiative conversion between spin isomers is not allowed in the gas
phase, preserving the OPR for long timescales. Gas-phase formation of
water occurs through exothermic reactions leading to an OPR of 3. On
grains, water forms and survives at low temperatures, and it is
tempting to equate $T_{\rm spin}$ with the grain temperature. However,
the energetics of water formation and ortho-to-para exchange on grains
are poorly understood\cite{limbach2006}, and the water OPR may be
changed by photodesorption. This process starts by dissociating water
to H and OH in the ice, and continues with the energetic H kicking out
a neighboring H$_2$O molecule from the ice matrix, or with the H and
OH recombining in the ice to form H$_2$O with sufficient internal
energy to sublimate\cite{andersson2008}.  The latter route drives the
OPR to at least unity, implying an even lower original ice OPR, to
yield a resulting OPR of 0.77.  Cometary volatiles are released
through thermal sublimation and their measured OPRs are interpreted to
reflect the OPR of their ice constituents. Equating $T_{\rm spin}$
with the physical temperature of the grain on which the ice formed is
supported by the similarity of measured $T_{\rm spin}$ of NH$_3$ and
H$_2$O in several individual solar system comets\cite{shinnaka2011}.

Solar system comets consist of a heterogeneous mixture of ices and
solids, likely assembled in the giant planet formation zone by mixing
local material with material that drifted in from larger
radii\cite{weidenschilling1977}.  Our water vapor observations probe
cold, ice-coated precometary grains residing beyond $>$50 AU
representing the bulk of the latter material.  The presence in comets
of crystalline silicates, requiring formation temperatures $>$800
K\cite{sandford2006}, together with CO and H$_2$O ices that condense
at 20 to 100 K, argues for transport of hot material from near the star
to the icy outer regions of the solar
nebula\cite{wooden2008}. Provided that spin temperatures reflect
formation histories, the different $T_{\rm spin}$ inferred for the
water ice in TW Hya ($<$13 K) and solar system comets ($>$20 K)
indicates a similar mixing of volatiles throughout the entire solar
nebula, blending water formed at $>$50 K and an OPR of 3, with water
formed at 10 to 20 K and OPR$<$1 probed by our observations. In this
case, the range of $T_{\rm spin}$ values of the cometary inventory
reflects the stochastic nature of transport and mixing.

Our Herschel detection of cold water vapor in the outer disk of TW Hya
demonstrates the presence of a considerable reservoir of water ice in
this protoplanetary disk, sufficient to form several thousand Earth
oceans worth of icy bodies. Our observations only directly trace the
tip of the iceberg of 0.005 Earth oceans in the form of water vapor.

% Your references go at the end of the main text, and before the
% figures.  For this document we've used BibTeX, the .bib file
% scibib.bib, and the .bst file Science.bst.  The package scicite.sty
% was included to format the reference numbers according to *Science*
% style.

%%%\bibliography{hogerheijde_twhya}
%%%\bibliographystyle{Science}

% Following is a new environment, {scilastnote}, that's defined in the
% preamble and that allows authors to add a reference at the end of the
% list that's not signaled in the text; such references are used in
% *Science* for acknowledgments of funding, help, etc.

%\begin{scilastnote}
%\item This work was partially supported by NWO grant 639.042.404, NSF
%   grant 0707777, and, as part of the NASA Herschel HIFI guaranteed
%  time program, by NASA. The presented data are archived at the
%  Herschel Science Archive, http://archives.esac.esa.int/hda/ui/,
%  under OBSID 1342198337 and 1342201585. 
%\end{scilastnote}

%%% Include acknowledgement without \scilastnote to suppress note #.

\noindent
Herschel is an European Space Agency space observatory with science
instruments provided by European-led principal investigator consortia
and with important participation from NASA.  This work was partially
supported by Nederlandse Organisatie voor Wetenschappelijk Onderzoek
grant 639.042.404, NSF grant 0707777 and, as part of the NASA Herschel
HIFI guaranteed time program, by NASA. The data presented here are
archived at the Herschel Science Archive,
http://archives.esac.esa.int/hda/ui/, under OBSID 1342198337 and
1342201585.

\bigskip
{\parindent 0pt
{\bf Supporting Online Material}\par
Materials and Methods\par
Table S1\par
References ({\it 28--39\/})
}

\clearpage
\clearpage

\begin{figure}
\includegraphics{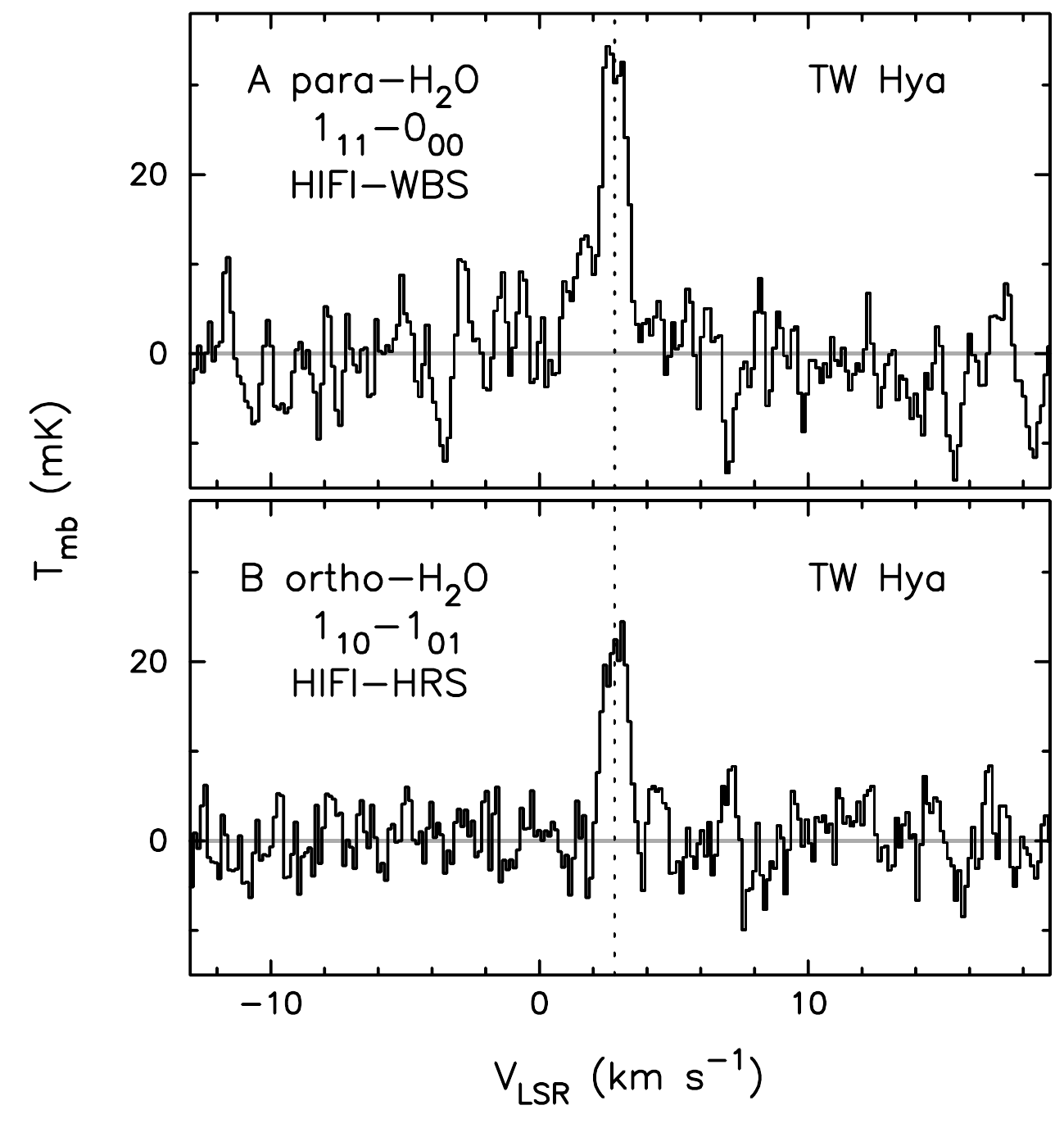}
\caption{Spectra of para-H$_2$O $1_{11}$--$0_{00}$
({\bf A}) and ortho-H$_2$O $1_{10}$--$1_{01}$ ({\bf B}) obtained with
HIFI on the Herschel Space Observatory toward the
protoplanetary disk around TW~Hya after subtraction of the continuum
emission. The vertical dotted lines show the system's velocity of +2.8
km~s$^{-1}$ relative to the Sun's local environment (local standard of
rest).}
\end{figure}

\clearpage

\begin{figure}
\includegraphics[width=16cm]{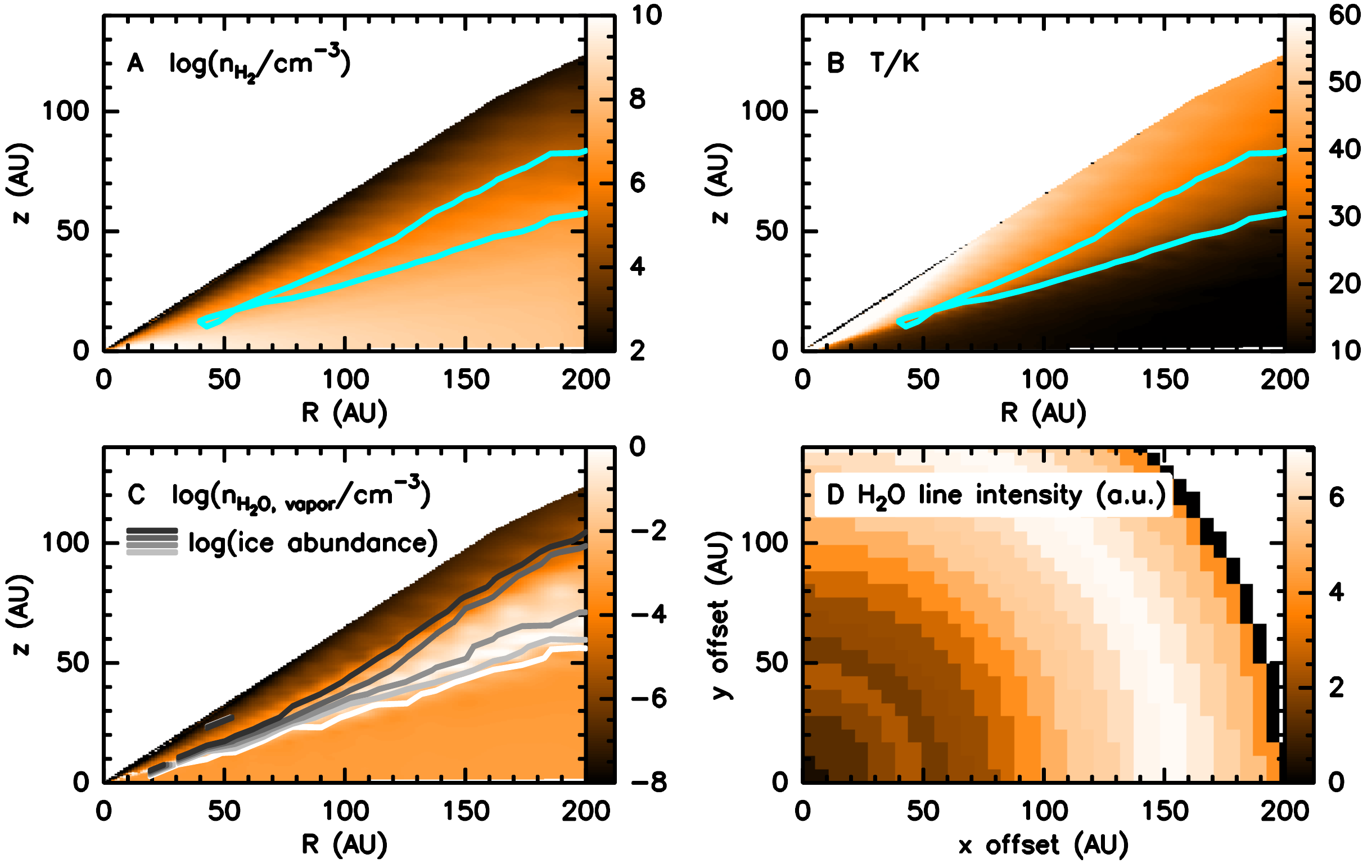}
\caption{ Adopted model for the TW Hya protoplanetary
disk: ({\bf A}) H$_2$ number density, ({\bf B}) dust temperature ,
({\bf C}) the number density of water vapor molecules and contours of
volume-averaged water ice abundance decreasing from white to black as
$2\times 10^{-4}$, $2\times 10^{-5}$, $2\times 10^{-6}$, $2\times
10^{-7}$, and $2\times 10^{-8}$, relative to H$_2$, and ({\bf D}) one
quadrant of the resulting water emission line intensity from the near
face-on disk, in arbitrary units. In (A) and (B), the blue contour
delineates the layer of maximum water vapor abundance.}
\end{figure}

% For your review copy (i.e., the file you initially send in for
% evaluation), you can use the {figure} environment and the
% \includegraphics command to stream your figures into the text, placing
% all figures at the end.  For the final, revised manuscript for
% acceptance and production, however, PostScript or other graphics
% should not be streamed into your compliled file.  Instead, set
% captions as simple paragraphs (with a \noindent tag), setting them
% off from the rest of the text with a \clearpage as shown  below, and
% submit figures as separate files according to the Art Department's
% instructions.

%\noindent {\bf Fig. 1.} Please do not use figure environments to set
%up your figures in the final (post-peer-review) draft, do not include graphics in your
%source code, and do not cite figures in the text using \LaTeX\
%\verb+\ref+ commands.  Instead, simply refer to the figure numbers in
%the text per {\it Science\/} style, and include the list of captions at
%the end of the document, coded as ordinary paragraphs as shown in the
%\texttt{scifile.tex} template file.  Your actual figure files should
%be submitted separately.

%----------------------------------------------------------------------

\clearpage
\pagestyle{plain}
\pagenumbering{arabic}
\baselineskip16pt
\section*{Supporting Online Material}

\section*{Materials and Methods}

\section{Observations and data reduction}

We observed TW Hya ($\alpha_{2000}=11^{\rm h}01^{\rm m}51.91^{\rm s}$,
$\delta_{2000}=-34^\circ 42' 17.0''$) with the Heterodyne Instrument
for the Far-Infrared (HIFI) onboard the Herschel Space
  Observatory in double beamswitch mode with a throw of $3.0'$, as
part of the `Water in Star-Forming Regions with Herschel' (WISH) key
program\cite{vandishoeck2011}. On 2010 June 15 the instrument observed
the H$_2$O $1_{10}$--$1_{01}$ line at 556.936002 GHz using receiver
band 1b and a local oscillator tuning of 551.895 GHz (OBSID
1342198337). A total on-source integration time of 181 min was
obtained and system temperatures were 75--95 K. On 2010 July 17 the
instrument observed the H$_2$O $1_{11}$--$0_{00}$ line at 1113.342964
GHz with receiver band 4b and a Local Oscillator tuning of 1108.221
GHz (OBSID 1342201585). A total on-source integration of 326 min was
obtained with system temperatures of 360--400 K.

The data were recorded in the Wide-Band Spectrometer (WBS) which
covers 4.4 GHz with 1.1 MHz resolution. At the observing frequencies
of 557 and 1113 GHz this corresponds to 0.59 and 0.30 km s$^{-1}$,
respectively. Simultaneously, the data were recorded in the
High-Resolution Spectrometer (HRS) which covers 230 MHz at a
resolution of 0.25 MHz (0.13 and 0.067 km s$^{-1}$ at the observing
frequencies of 557 and 1113 GHz, respectively). Both horizontal and
vertical linear polarizations were measured. The relative offsets of
% (*)
the respective polarization channels from our pointing center were
$3.3''$ at 556 GHz and $1.8''$ at 1113 GHz, in both cases much smaller
than the beams of $38.1''$ and $19.1''$, respectively, and not
relevant for our analysis.

The in-orbit calibration system converted the raw data to the $T_{\rm
  A}^*$ (antenna temperature) intensity scale. Further conversion to
the main-beam antenna temperature scale $T_{\rm mb}$ was done using a
beam efficiency of 0.76 at both observing frequencies, accurate to
10\%. The data were reduced using HIPE v4.0.0 (1113 GHz) and v 4.6.0
(556 GHz), and exported to the CLASS software package for further
analysis. The velocity accuracy of the final data product is better
than a few m~s$^{-1}$.

After averaging together the horizontal- and vertical-polarization
measurements, resulting noise levels are 1.6 mK per 0.27 km s$^{-1}$
WBS channel at 557 GHz, 4.2 mK per 0.065 km s$^{-1}$ HRS channel at
557 GHz, 4.95 mK per 0.135 km s$^{-1}$ WBS channel at 1113 GHz, and
14.3 mK per 0.032 km s$^{-1}$ HRS channel at 1113 GHz. Throughout the
long integration, the noise was confirmed to decrease proportional to
(time)$^{-0.5}$\cite{bergin2010}.

Linear baselines were fit using a 50 km~s$^{-1}$ interval centered on
the source velocity, but excluding a 5 km~s$^{-1}$ interval around the
emission lines. This yielded continuum levels of $5.60\pm 0.05$ at 557
GHz and $13.2\pm 0.1$ Jy at 1113 GHz, which compare well with measured
fluxes in the literature\cite{qi2006}. After subtraction of the
continuum level, single Gaussian profiles were fitted to the clearly
visible lines yielding values for the Local Standard of Rest Velocity,
$V_{\rm LSR}$, of the lines, their Full Width at Half Maximum, their
peak intensity and their integrated intensity (Table S1).

\section{Physical and chemical model of the TW Hya disk}

The adopted physical model for the TW Hya disk is based on recent
modeling of its full spectral energy distribution including Spitzer
and Herschel data\cite{thi2010}, with parameters similar to
those that reproduce a wide set of observations including gas emission
lines\cite{gorti2011}. It has a dust mass of $1.9\times 10^{-4}$
M$_\odot$ in $<1$~mm particles, and a gas mass of $1.9\times 10^{-2}$
M$_\odot$ obtained by assuming a gas-to-dust mass ratio of 100.  The
adopted disk dust mass is a
%fiducial (27)
lower limit, containing only particles $<$1~mm and employing a
close-to maximum emissivity to turn continuum flux into mass: other
mass estimates are four times larger (in $<$1 mm particles), or as
much as 14 times larger if significant grain growth (as expected) is
included and 93\% of the mass is $>1$ mm\cite{thi2010}. The gas mass
is near the high end of the inferred gas masses in the literature of
$4\times 10^{-5}$--0.06
M$_\odot$\cite{kastner1997,thi2010,gorti2011}. This range reflects the
uncertainty in translating observations into gas-mass estimates. The
low end of the range corresponds to taking CO observations, assuming
an abundance of $10^{-4}$ with respect to H$_2$, and ignoring any
freeze out\cite{kastner1997}. The high end is derived from dust
continuum observations and a standard gas-to-dust ratio of 100
\cite{calvet2002}, but is found to reproduce gas-line observations
including CO when more realistic chemical conditions are
adopted\cite{gorti2011}. We will explore the effects of gas masses
lower than the 0.02 M$_\odot$ from our baseline model.

The disk's outer radius is 196 AU, and its surface density outside a
radius of 4 AU is given by $\Sigma(R) = 1.4\times 10^{-2}\,
(R/100\,{\rm AU})^{-1.0}$ g~cm$^{-2}$. Inside 4 AU the surface density
is much lower, but we can ignore this region in our modeling. The
vertical density distribution is characterized by an exponential scale
height $H(R)=10\,(R/100\,{\rm AU})^{1.2}$ AU and a density profile
$\rho (z,R) \propto \exp(-z^2/2H^2)$. The dust particles have a size
distribution $N(a) \propto a^{-3.4}$ between $a_{\rm min}$=0.03 $\mu$m
and $a_{\rm max}$=10~cm, and a mean density of 3.5 g~cm$^{-3}$. Gas
and dust are assumed to be well mixed throughout the disk. The dust
temperature is calculated for a stellar effective temperature of 4000
K and a luminosity of 0.23 L$_\odot$ with the RADMC
code\cite{dullemond2004}, which yields a result identical to the
original model. The gas and dust temperatures are assumed to be
identical.

The penetration of the stellar ultraviolet radiation, both continuum
and Ly$\alpha$, is calculated for the defined density structure,
adopting the measured stellar spectrum and ultraviolet dust scattering
properties from\cite{weingartner2001}. Subsequently, the chemical
composition throughout the disk is calculated following the methods of
\cite{fogel2011} including the effects of gas-phase reactions, thermal
desorption, desorption by ultraviolet radiation\cite{oberg2009},
cosmic-ray induced ultraviolet photons, and photodissociation by
ultraviolet radiation. X-ray propagation and dissociation of H$_2$ by
other species is also included in the calculation, but do not
influence our results.
% (*)
Cosmic ray induced photodesorption provides
the only source of water vapor in the disk interior, leading to a low
base-level of water vapor abundance of $\sim 10^{-10}$ relative to
H$_2$\cite{shen2004}. Our simulations do not treat the disk structure
and water chemistry in the inner disk ($<5$ AU) correctly, but in the
large Herschel beam these regions contribute only negligible
amounts of line flux.  Except for the inner several AU, the
temperatures in the disk are everywhere well below 200 K, excluding
the efficient formation of water in the gas phase, leaving
photodesorption of water ice as the sole source of water vapor to
explain the Herschel observations.

This baseline model contains a total water ice
% (5)
reservoir of 6300 Earth Oceans (equivalent with $9\times 10^{27}$ g,
with one Earth Ocean containing $1.5\times 10^{24}$ g of water),
mostly as ice frozen out onto dust grains. This ice reservoir is a
rough estimate at best. It is based on a low dust mass of $1.9\times
10^{-4}$ M$_\odot$. It assumes that all water ice thought to be
present in the protostellar core from which TW Hya formed was either
retained or efficiently reformed from gas-phase oxygen on the dust
grains. We adopt an oxygen abundance of $3.5\times 10^{-4}$ relative
to H, with 70\% locked up in water and 30\% in CO\cite{visser2009}. As
such, our model may overestimate the ice reservoir if the disk is
formed depleted in elemental oxygen, because the amount of water ice
scales linearly with the oxygen abundance.
% (19)
This may be the case, if the water ice formed after the disk underwent
significant loss of its gas mass. For example, if the disk lost 90\%
of its gas mass before the water ice formed, the water ice reservoir
would be lowered by a corresponding factor.

\section{Excitation of the water molecule and formation of the
  emission lines}

Most water vapor is present at H$_2$ number densities of a few times
$10^6$ cm$^{-3}$ and temperatures of 25--35 K. This necessitates a
full calculation of the statistical equilibrium populations of the
involved levels, because these values are below the critical density
and upper level energies for the transitions. The column density of
water molecules, and resulting line opacities, further necessitates a
full calculation of the transport of line photons including absorption
and resonant scattering. We calculated the statistical equilibrium
excitation and emission line formation with the LIME
code\cite{brinch2010}. Only collisions with H$_2$ are included; at a
relative fraction of $<10^{-7}$ relative to H$_2$ in the region of
interest, electrons do not contribute significantly to the
excitation. Collisional rate coefficients were taken from
\cite{faure2007,dubernet2009} and an H$_2$ ortho-to-para ratio of 3.0
was adopted. The water vapor ortho-to-para ratio in our standard model
is characterized by a spin temperature equal to the local dust
temperature.

Our simulations show that the excitation of the relevant energy levels
is subthermal, as expected for densities lower than the critical
densities of the transitions of $\sim 2\times 10^7$ cm$^{-3}$ for
H$_2$O $1_{10}$--$1_{01}$ and $\sim 2\times 10^8$ cm$^{-3}$ for H$_2$O
$1_{11}$--$0_{00}$. In the standard model the mean free path of the
photons is small compared to the thickness of the layer with abundant
water vapor. Still, the emission is effectively optically thin,
because of the significant chance of resonant scattering compared to
photon absorption in this excitation regime.  Crucially, this means
that our observations are sensitive to the total amount of water vapor
as opposed to fully optically thick lines. However, it also means that
the calculated emission of the individual lines depends sensitively on
the location of the peak water abundance, since from this follows the
surrounding H$_2$ number density and therefore the excitation
conditions. The sensitivity of the excitation on the gas temperature
is much smaller.

% EAB this is a good point maybe it can be put back in....
%At the wavelengths of the lines, the
%disk's dust is optically thin outside radii of 3--10 AU.   

% (20)
Our standard model overestimates the water fluxes by factors 3--5
compared to the observations. The line strengths can be reduced if the
collisional excitation is decreased or if the total amount of water
vapor is lowered. We first explore the dependency of our results on
the collisional excitation, through the effects of the overall H$_2$
density and the H$_2$ ortho-to-para ratio (OPR). The results of our
standard model do not change significantly when we increase the H$_2$
density or change the H$_2$ OPR from 3 to 0.1. The latter corresponds
to an H$_2$ OPR in thermal equilibrium at a temperature of $\sim 35$ K
as appropriate for the layer of maximum water abundance, and
represents a lower limit to the H$_2$ OPR. The lack of dependency can
be explained by the small mean free path of the water line photons in
the standard model, which contributes to the excitation of the energy
levels. Only when the H$_2$ density is lowered by an order of
magnitude does the excitation of the water molecules and the emergent
line strengths decrease. Within our model, decreasing the H$_2$
density corresponds to increasing the height of the layer of maximum
water abundance. Since the latter is set by the vertical distribution
of the dust, this corresponds to increasing the height of the
dust. However, the dust is supported vertically by the gas pressure,
and increasing the vertical height of the dust is therefore not
allowed. Decreasing the height of the dust is allowed, as happens when
all grains, independent of grain size, decouple partially from the gas
and settle uniformly to the midplane. This moves the layer of maximum
water abundance to a region of increased H$_2$ density, but does not
affect the line strengths significantly because the increased
excitation is balanced by increased line opacity.

% (4)
Reducing the total gas mass of the disk does not decrease the strength
of the water emission lines. The reason for this is two-fold. First,
we assume that the water ice, from which the water vapor derives
through photodesorption, was formed on the grains early in the disk's
history, before any significant gas loss occurred, and remains
unchanged. Therefore, even for low disk gas masses, the column density
of water vapor is unchanged. Secondly, as the gas density drops, the
ice-carrying grains, which are vertically supported by the gas
pressure, must settle to the midplane. As a result, we expect the
H$_2$ density in the layer of water vapor to remain similar as
the vertical height of the layer is reduced. The collisional
excitation of the water vapor molecules therefore is not changed
significantly. Together, these two effects lead to essentially
unchanged water vapor emission line strengths and unchanged total
amount of ice even if the gas-to-dust ratio is lowered. This reasoning
rests crucially on the assumption that the water ice formed at an
original gas-to-dust ratio of 100. If the ice formed after 90\% of the
gas has been lost, the resulting ten times smaller ice reservoir can
explain our observations. However, there are no likely scenarios in
which the water ice starts to form only at such an advanced stage of
the disk's evolution.

The lack of viable scenarios to reduce the intensities of the water
lines by changing the H$_2$ density, its OPR, or the total gas mass
leaves as our only option a decrease of the total amount of water
vapor to match the observations. Since the water vapor derives from
photodesorption of ice, this means that the amount of water ice in
the regions affected by ultraviolet radiation needs to be reduced. We
calculated models where the abundance of ortho-water and para-water
were reduced uniformly throughout the disk with separate factors until
the emerging line intensities match the observations. Unlike our
standard model, $T_{\rm spin}$ now no longer is assumed equal to the
dust temperature but instead is constant throughout the disk. Compared
to the total amount of water vapor predicted by the standard model, a
model where $5.4\pm 0.2$\% is present as ortho-water and $7.0\pm
0.3$\% as para-water reproduced the observed line intensities (i.e., a
water OPR of $0.77\pm 0.07$). The error bars reflect only the
statistical uncertainty due to the measurement error in the observed
line intensities; systematic uncertainties are discussed below. This
corresponds to a reduction to 12\% of the ice content in the regions
affected by the ultraviolet radiation, although the size of the total
ice reservoir does not change. At these low abundances, the disk
becomes fully optically thin to the water line photons.

We finally investigated the dependency of the derived water OPR at
these reduced abundances on the water excitation through collisions
with H$_2$, including the effect of the H$_2$ OPR. Because the
critical density of the para-H$_2$O $1_{11}$--$0_{00}$ line is higher
than that of the ortho-H$_2$O $1_{10}$--$1_{01}$ line, increasing the
H$_2$ density increases the para-H$_2$O line faster than the
ortho-H$_2$O line, which decreases the water OPR derived from the
observed line ratio. A factor of 10 increase in the H$_2$ density
increases the derived water OPR to 2.0. Such an increase in H$_2$
density corresponds to significant uniform settling of all dust, both
large and small, and the associated layer of maximum water vapor
abundance, for which there is no observational evidence in the TW Hya
disk at the relevant radii. It would furthermore require an even
larger degree of differential settling of large, icy grains with
respect to small grains to further reduce the water vapor abundance in
order to reproduce the observed line strengths. No degree of uniform
settling can yield a derived water OPR of 3.

In the optically thin limit appropriate for the disk model with
reduced water vapor abundance, reducing the H$_2$ OPR from 3 to 0.1
has an effect comparable to lowering the H$_2$ density since
para-H$_2$ has lower collision rates with water compared to
ortho-H$_2$. Because of the differences in collision rates between
ortho-water and para-water, the derived water OPR increases to 1.3 for
an H$_2$ OPR of 0.1. Thus, our conclusion is robust that the OPR in
the TW Hya disk is much lower than the range of 2--3 measured for
Solar System comets.

From these considerations we conclude that the most likely explanation
of our Herschel detection of the emission lines of water vapor
from the TW Hya disk, is a situation where only 12\% of the original
ice content remains in the regions affected by UV irradiation and the
remainder has settled to lower depths. The ortho-to-para ratio of the
water vapor, and by extension the ice reservoir, is as low as 0.77 to
reproduce the observed ratio of the $1_{11}$--$0_{00}$ and
$1_{10}$--$1_{01}$ lines.

\bigskip
\bigskip
\bigskip

\begin{tabular}[t]{lcccc}
\multicolumn{5}{c}{\hfil\bf Table S1. Observed line parameters.}\\
\hline
Transition & $V_{\rm LSR}$ & FWHM & $\int T_{\rm mb}dV$  & $F_{\rm line}$\\
 & (km s$^{-1}$) & (km s$^{-1}$) & (mK km s$^{-1}$) & ($10^{-19}$ W m$^{-2}$)\\
\hline
H$_2$O $1_{11}$--$0_{00}$ & 
   $2.73\pm 0.08$ & $1.2\pm 0.2$ & $44.6\pm 2.8$ & $3.07\pm 0.19$\\
H$_2$O $1_{10}$--$1_{01}$ & 
   $2.84\pm 0.04$ & $0.96\pm 0.07$ & $25.2\pm 1.1$ & $3.46\pm 0.15$ \\
\hline
\end{tabular}

\bigskip
\bigskip
\bigskip

%\bibliography{hogerheijde_twhya}

\begin{thebibliography}{10}
\bibitem{aunote}
One astronomical unit (AU) is the mean distance between Earth and the Sun
  of $1.49598\times 10^{11}$ m.
\end{thebibliography}
%\bibliographystyle{Science}
%\clearpage

\noindent
Herschel is an European Space Agency space observatory with science
instruments provided by European-led principal investigator consortia
and with important participation from NASA.This work was partially
supported by Nederlandse Organisatie voor Wetenschappelijk Onderzoek
grant 639.042.404, NSF grant 0707777 and, as part of the NASA
Herschel HIFI guaranteed time program, by NASA. The data presented
here are archived at the Herschel Science Archive,
http://archives.esac.esa.int/hda/ui/, under OBSID 1342198337 and
1342201585.

\end{document}